\documentstyle[12pt]{article}
\newskip\humongous \humongous=0pt plus 1000pt minus 1000pt

\newif\ifdtup

\def\tr{\mathop{\rm tr}}

\def\beq{\begin{equation}}
\def\eeq{\end{equation}}

\def\beqn{\begin{eqnarray}}
\def\eeqn{\end{eqnarray}}
\relax

\def\G2{{\; \rm GeV/}c^2}
\def\G{\; \rm GeV}

\def\dotx{\dotx{\dot\overline{x}}}

\relax

\begin{document}
\begin{titlepage}
\begin{flushright}
       {\normalsize    OU-HET 236   \\  hep-th 9604***\\
                April,~1996  \\}
\end{flushright}
\vfill
\begin{center}
  {\large \bf     Schwinger-Dyson and Large $N_{c}$ Loop Equation for
      \\  Supersymmetric Yang-Mills Theory }\footnote{This work is
 supported in part by  Grant-in-Aid for  Scientific Research
(07640403)
from
 the Ministry of Education, Science and Culture, Japan.}

\vfill

         {\bf H.~Itoyama}  \\
            and \\
              {\bf H.~Takashino}
\vfill
        Department of Physics,\\
        Faculty of Science, Osaka University,\\
        Toyonaka, Osaka, 560 Japan\\
\end{center}
\vfill
\begin{abstract}
    We derive  an infinite sequence of  Schwinger-Dyson equations  for
  $N=1$ supersymmetric Yang-Mills theory.
  The fundamental and the only variable employed is the Wilson-loop
  geometrically represented in $N=1$ superspace:
 it organizes an infinite number of supersymmetrizing insertions
 into  the ordinary Wilson-loop as a single entity.
 In the large $N_{c}$ limit,  our equation becomes
  a closed loop equation for the one-point function of the Wilson-loop
  average. 
\end{abstract}
\vfill
\end{titlepage}

  A number of recent developments on nonperturbative phenomena associated with
  supersymmetric gauge theories suggest that old questions such as quark
 confinement, monopole condensation and dynamical symmetry breaking  can be
 understood  better  when supersymmetry is operating. In years to come,
 these developments  may even be extended to physics of superstrings to which
 nonperturbative  formulation is still largely lacking.

  Schwinger-Dyson approach  is a formulation which has been successful
  in deriving exact results in matrix models  describing  toy models
  of noncritical strings. It has supplied some valuable results in more
 realistic theories with manageable approximations. 
  Clearly the most transparent formulation  of the Schwinger-Dyson equations
 in gauge theories is given in terms of the gauge invariant Wilson-loop
 variable.
 It is well-known that one can write down the infinite sequence
 of  equations  using the  geometric operations alone acting on the loop
 in pure Yang-Mills theory.
 In the large $N_{c}$ limit, it reduces to a closed loop equation
  for the one-point function of the Wilson-loop average \cite{MM}. In this
 letter, we will provide its $N=1$ supersymmetric counterpart.\footnote{Matter
  supermultiplets can be added to produce another term in the Schwinger-Dyson
  equations.}

  Despite its obvious interest and relevance to the motivations decades ago
 and  some current issues, this problem we
 will address ourselves below has neither been thoroughly investigated nor
 completed in the literature. 
   See  \cite{GJ}, \cite{MakMed}  for earlier references.
   The fundamental variable  we will employ is
  the gauge invariant Wilson-loop which  is defined in terms of 
 the path-ordered exponential in $N=1$ superspace
 \cite{Gates},\cite{Mar},\cite{GGRS}.
 
  Let us briefly recall  the case of ordinary Yang-Mills theory:
 ${\cal L}_{YM} = \frac{(ig)^{2}}{4kg^{2} } tr v_{mn} v^{mn}~,$
  $~ig v_{mn} \equiv [D_{m}, D_{n}]~,$
$D_{m} \equiv  \partial_{m} +ig v_{m}~,$ $~v_{m} \equiv v_{m}^{(r)} T^{r}~,$
$tr T^{r}T^{s} = k \delta^{rs}$.   We take the gauge group to be
 $U(N_{c})$  for simplicity. 
 We denote by ${\cal C}_{xy}$ a path which
  begins with $x$ and ends at $y$.
    Let  ${\bf W}[{\cal C}] \equiv  <<\frac{k}{N_{c}} tr U_{ {\cal C}}>>$
  with  $ U_{ {\cal C}} \equiv  P\exp ( -ig \oint_{ {\cal C} }
 dz^{m} v_{m}(z) )$  be the one-point function of the Wilson-loop average
  associated with any closed loop ${\cal C}$  in Minkowski space.  
 A fundamental role is played by the area derivative  acting on
  the Wilson-loop, which produces the field strength.
  The subsequent action of the ordinary derivative supplies the equation
  of motion of Yang-Mills theory:
\beqn
\label{eq:(1)}
\frac{\partial}{\partial x_{\ell}} \frac{\partial}{\partial
 \Sigma^{m \ell}(x) } {\bf W}[{\cal C}] = -ig  <<\frac{k}{N_{c}} tr D^{\ell}
 v_{m \ell} (x) U_{ {\cal C}_{xx}}>> \;\;.
\eeqn
  We remind the readers that the area derivative acting on a functional
 $f[ {\cal C}_{yz}]$
 can be defined through an increment associated with an addition of 
 an infinitesimal area element at the point $x^{m}= x^{m}(s,t)$ on the path
 ${\cal C}_{yz}$:
\beqn
\label{eq:aread}
 \Delta f[{\cal C}_{yz}] &\equiv& f[{\cal C}_{yz} + \delta {\cal C}_{x}^{(1)}]
 - f[ {\cal C}_{yz} + \delta {\cal C}_{x}^{(2)} ] \;\;\; \nonumber \\
  &\equiv& \frac{\partial f[{\cal C}_{yz}]}{ \partial \Sigma^{m \ell}(x)}
  \frac{1}{2} \left( \frac{\partial x^{m}}{ \partial s} \frac{ \partial 
 x^{\ell}}{ \partial t} - \frac{\partial x^{m}}{ \partial t} \frac{ \partial
 x^{\ell}}{ \partial s} \right) \Delta s \Delta t \;\;\; \nonumber \\
 &\equiv&  \frac{\partial f[{\cal C}_{yz}]}{ \partial \Sigma^{m \ell}(x)}
  \left(\Delta  \Sigma  \right)^{m \ell}(x; \Delta s, \Delta t) \;\;\;.
\eeqn
   Here $\delta {\cal C}_{x}^{(1)}$ is a path consisting of two straight lines
 connecting $x^{m}(s,t)$ with $x^{m}(s +\Delta s, t)$ and
 $x^{m}(s +\Delta s, t)$  with $x^{m}(s +\Delta s, t + \Delta t)$.  Similarly,
 $\delta {\cal C}_{x}^{(2)}$  is obtained by replacing $x^{m}(s +\Delta s, t)$
 by $x^{m}(s, t +\Delta t)$ in the above.
 This definition can be extended to superspace in a straightforward fashion.

 The right hand side of eq.~(\ref{eq:(1)}) can of course be written as
\beqn 
\label{eq:step}
 g  \int [{\cal D}v_{m}]  \frac{k}{N_{c}} tr T^{r}
 (\frac{\delta e^{i S_{YM}}}{\delta v^{(r)m}(x)})  U_{{\cal C}_{xx}}  \;\;.
\eeqn
  Upon partial integration  followed by the use of the completeness relation,
 eq.~(\ref{eq:step})  produces an  equation for the one- and two-loop
 correlators, which is  the first among the infinite sequence
  of the Schwinger-Dyson equations:
\beqn
\label{eq:SD12}
\frac{\partial}{\partial x_{\ell}} \frac{\partial}{\partial
 \Sigma^{m\ell}(x) } {\bf W}[{\cal C}] =   ig^{2}
   \oint_{\cal C}   dz_{m} << \frac{k^{2}}{N_{c}}
\left(\tr U_{{\cal C}_{zx}} \right) \delta^{(4)}(x-z)
 \left(\tr U_{{\cal C}_{xz}} \right) >> .
\eeqn
  Obviously one can generate the rest of the equations in the same way.
  We find
\beqn
&&\frac{\partial}{\partial x_{\ell}} \frac{\partial}{\partial
 \Sigma^{m\ell}(x) } << \left( \frac{k}{N_{c}} \right)^{n}
    \prod_{i=1}^{n} tr U_{ {\cal C}^{(i)} } >>  \nonumber \\
 &=&   ig^{2} N_{c}
   \sum_{j=1}^{n} \Theta ( x \in {\cal C}^{(j)})\oint_{{\cal C}^{(j)} } 
  dz_{m} <<  \left( \frac{k}{N_{c}} \right)^{n+1}
 \left( \prod_{i=j+1}^{n} tr U_{ {\cal C}^{(i)} } \right)
 \left(\tr U_{{\cal C}_{zx}^{(j)}} \right)  \nonumber \\   
 && \delta^{(4)}(x-z)
 \left(\tr U_{{\cal C}_{xz}^{(j)} } \right) 
  \left( \prod_{i=1}^{j-1} tr U_{ {\cal C}^{(i)} } \right) >> \;.
\eeqn
  Here  $ \Theta ( x \in {\cal C})$ is $1$  when $x$ belongs to
  the loop ${\cal C}$ and $0$ otherwise. 
  In the large $N_{c}$ limit   with $\lambda_{c} \equiv  g^{2} N_{c}$
  kept finite, eq.~(\ref{eq:SD12}) reduces to the closed loop equation for
 $W[{\cal C}]$
  albeit being highly singular \cite{MM}:
\beqn
\frac{\partial}{\partial x_{\ell}} \frac{\partial}{\partial \Sigma^{m \ell}
(x) } {\bf W} [{\cal C}] = i \lambda_{c}  \oint_{\cal C} 
  dz_{m}  {\bf W}[{\cal C}_{zx}]    \delta^{(4)}(x-z)
{\bf W}[{\cal C}_{xz}] \;\;\;.
\eeqn 

 We find   that  the above reasoning  has a straightforward  generalization
 to $N=1$ supersymmetric Yang-Mills theory 
  in its geometric formulation on $N=1$ superspace.\footnote{The above 
derivation can, of course, be presented as
 $ 0 = \int [{\cal D}v_{m}]  tr \left( T^{r}\right.$
 $\left. 
\frac{\delta}{\delta v_{m}^{(r)}(x)} \left( U_{{\cal C}} e^{i S_{YM}}
  \right) \right)$.
   This, however, does not lead to a fruitful superspace generalization.}
  We use the same symbols  for   the objects appearing above   and
   their  superspace counterparts.  We closely follow the notation of
 \cite{WB}.  The action for  the supersymmetric Yang-Mills  theory reads
\beqn
   S_{SYM}  = \int d^{4}x  \frac{1}{8kg^{2}} tr ( -2i\lambda \sigma^{a}
   D_{a} \bar{\lambda} + D^{2} - \frac{1}{2} v_{ab}v^{ab} ) \;\;\;.
\eeqn                        
   This time,  let   the supersymmetric Wilson-loop associated with a closed
 loop ${\cal C}$ in superspace \cite{Gates},\cite{Mar},\cite{GGRS} be
\beqn
  {\bf  W}[{\cal C}] \equiv  <<\frac{k}{N_{c}} tr U_{ {\cal C}}>> \nonumber \\
    {\rm with}~~~  U_{ {\cal C}} \equiv  P\exp ( \oint_{ {\cal C} }
  \phi ) \;\;,
\eeqn
  where   $\phi \equiv dz^{M}\phi_{M}(z) = dz^{M}\phi_{M}^{(r)}(z)iT^{r}$
  is  the connection one-form   which is  Lie algebra valued.\footnote{
  A rescaling by factor $2g$  in the  gauge potentials from the
 non-supersymmetric case  is understood and $\phi_{M}$ is antihermitean.}
    A solution to the  proper set of  constraints of flat connections
  on superspace    together with the constraints imposed  by
   a set  of  Bianchi identities provides  an expression in terms of
   the real superfield $V(x,\theta, \bar{\theta})$  ( see for instance
 \cite{So},\cite {WB}):
\beqn 
 (-2){\cal F}_{a \dot{\alpha} } &=& -i W^{\beta}
 \sigma_{a \beta \dot{\alpha} } \;\;, \\
 \phi_{\alpha} &=& - e^{-V} D_{\alpha} e^{V} \;\;, \;\; 
\phi_{\dot{\alpha}} =0 \;\;, \;\;
 \phi_{a} = \frac{i}{4} \bar{\sigma}_{a}^{ \dot{\alpha} \alpha}
  \bar{D}_{\dot{\alpha}} e^{-V} D_{\alpha} e^{V} \;\;, \\
  W_{\alpha} &=&   - \frac{1}{4}\bar{D}^{2} e ^{-V} D_{\alpha} e^{V} \;\;. 
\eeqn
  In  calculation  we employ  
 the $y$ coordinate
  ( $y^{a} \equiv x^{a}+i \theta \sigma^{a} \bar{\theta}$) \cite{WB}, in terms
 of which
\beq  
 \phi (y, \theta, \bar{\theta}) = d y^{a} \phi_{a}(y, \theta, \bar{\theta})
 -2i d \theta \sigma^{a} \bar{\theta} \phi_{a}(y, \theta, \bar{\theta})
  +  d \theta^{\alpha} \phi_{\alpha} (y, \theta, \bar{\theta}) \;\;.
\eeq

   Area derivative in superspace  can be constructed in an entirely
  analogous way as in  Minkowski space.
    For any functional $f[{\cal C}_{yz}]$  on superspace,  the area
 derivative $ \partial/\partial  \Sigma^{M L}(x)$   can be
 defined in any direction including the grassmannian ones by considering
\beqn
\label{eq:saread}
 \Delta f[{\cal C}_{yz}] &\equiv& f[{\cal C}_{yz} + \delta {\cal C}_{x}^{(1)}]
 - f[ {\cal C}_{yz} + \delta {\cal C}_{x}^{(2)} ] \;\;\; \nonumber \\
  &\equiv& \frac{\partial f[{\cal C}_{yz}]}{ \partial \Sigma^{M L}(x)}
  \frac{1}{2} \left( \frac{\partial x^{M}}{ \partial s} \frac{ \partial 
 x^{L}}{ \partial t} - \frac{\partial x^{M}}{ \partial t} \frac{ \partial
 x^{L}}{ \partial s} \right) \Delta s \Delta t \;\;\;.
\eeqn
 We find
\beqn
\label{eq:DWS} 
 tr (\sigma_{\alpha \dot{\beta}}^{a} \epsilon^{\dot{\beta} \dot{\alpha} }
   D^{\alpha} \frac{\partial U_{ {\cal C}_{yy} }  }
{\partial \Sigma^{a \dot{\alpha}} (y) })
  = 2i tr \left( ({\cal D}W)  U_{{\cal C}_{yy} } \right) \;\;,
\eeqn
  where
\beqn
 {\cal D} W \equiv DW -\phi W + W \phi  
   &=& D^{\alpha}W_{\alpha} - \{ \phi^{\alpha}, W_{\alpha} \} \\
  \frac{\partial U_{ {\cal C}_{yy} }  }
{\partial \Sigma^{A B}  (y) }  &=& F_{A B}(y) U_{ {\cal C}_{yy} } \;\;\;.
\eeqn
 Here $A,B$ are flat indices in superspace.

  In order to convert
 $\sigma_{\alpha \dot{\beta}}^{a} \epsilon^{\dot{\beta} \dot{\alpha} }
 D^{\alpha} \frac{\partial }{\partial \Sigma^{ a \dot{\alpha} } }
 {\bf W}[{\cal C}]$
  into another expression,   we compute ${\cal D} W$ in the W-Z gauge.
  For that purpose,  we first find
the expression of $\phi_{a}$ and that of $\phi_{\alpha}$ in the W-Z gauge: 
\beqn
  \xi\phi=\xi\sigma^{a}\bar{\theta}v_{a}-2i\xi\theta\bar{\theta}\bar{\lambda}
  +i\bar{\theta}\bar{\theta}\xi\lambda-\xi\theta\bar{\theta}\bar{\theta}D
  +i\bar{\theta}\bar{\theta}\xi\sigma^{ab}\theta v_{ab}-\theta\theta
  \bar{\theta}\bar{\theta}\xi\sigma^{a}D_{a}\bar{\lambda}, \\
2i\phi_{a}= v_{a}-i\bar{\lambda}\bar{\sigma_{a}}\theta
  +i\bar{\theta}\bar{\sigma_{a}}\lambda-\bar{\theta}\bar{\sigma_{a}}\theta D
  +i\bar{\theta}\bar{\sigma_{a}}\sigma^{bc}\theta v_{bc}
-\theta\theta\bar{\theta}\bar{\sigma_{a}}\sigma^{b} D_{b}\bar{\lambda}\;\;\;.
\eeqn
  These are again in  the $y$ coordinate. 
   After some amounts of algebras, we find 
\beqn
- {\cal D} W  &=& 2D + 2 \bar{\theta}\bar{\sigma}^{a} D_{a} \lambda
  -2 \theta \sigma^{a} D_{a} \bar{\lambda}
  + 2 \theta \sigma_{b}\bar{\theta} ( D_{a} v^{ab} -\frac{1}{2}
 \bar{\sigma}^{b \dot{\beta} \beta} \{ \bar{\lambda}_{\dot{\beta}} ,
 \lambda_{\beta} \}  )   \nonumber \\
 &-& 2i \theta \sigma^{a} \bar{\theta} D_{a} D +
 2i \theta \theta \bar{\theta} \bar{\sigma}^{b} \sigma^{a} D_{b} D_{a}
  \bar{\lambda} +2i \theta \theta [ D, \bar{\theta} \bar{\lambda}]\;\;,
\eeqn
  where $ D_{a} \cdots \equiv  [\partial_{a} + \frac{i}{2} v _{a},  \cdots]$. 
   As ${\cal D} W =0$ summarizes equations of motion for  $v_{m}, \lambda,
 \bar{\lambda}, D$, it should be that  ${\cal D} W $ is written as a
 particular variation of $S_{SYM}$  as well.
We find
\beqn
\label{eq:dvmod}
  &-& {\cal D} W =
 8g^{2}   \left \{  T^{r} \frac{\delta}{\delta D^{(r)}} +i T^{r}
 \bar{\theta}_{\dot{\alpha}} \frac{\delta}{\delta
 \bar{\lambda}_{\dot{\alpha}}^{(r)} }
 -i T^{r} \theta^{\alpha} \frac{\delta}{\delta \lambda^{\alpha (r)}}
    - T^{r} \bar{\theta}\bar{\sigma}_{a} \theta 
  \frac{\delta}{\delta v_{a}^{(r)}}  \;\;\;
 \right.   \nonumber \\
 &+& i\bar{\theta}\bar{\sigma}_{a} \theta D_{a} T^{r}
 \frac{\delta}{\delta D^{(r)}}
 - \theta \theta (\bar{\theta} \bar{\sigma}^{b})^{\alpha}
  D_{b}T^{r} \frac{\delta}{\delta \lambda^{\alpha (r)}}
 +i \theta \theta    \left.  [ T^{r} \frac{\delta}{\delta D^{(r)}},
  \bar{\theta} \bar{\lambda} ]  \right \}  S_{SYM} \;\;.
\eeqn
We denote this expression by
 $8g^{2} \delta  S_{SYM} / \delta V_{mod}(y, \theta,\bar{\theta})$.

  From eq.~(\ref{eq:DWS}) and eq.~(\ref{eq:dvmod}), we obtain
\beqn
\label{eq:ssd}
&&\sigma_{\alpha \dot{\beta}}^{a} \epsilon^{\dot{\beta} \dot{\alpha} }
   D^{\alpha} \frac{\partial }{\partial \Sigma^{ a \dot{\alpha}}
(y^{\prime})  } {\bf W } [{\cal C}]  \nonumber \\
  &=& - 16g^{2} \int[{\cal D} v_{m}][ {\cal D} \lambda ]
 [ {\cal D}\bar{\lambda} ] [ {\cal D} D] 
  \frac{k}{N_{c} }  tr  \frac{\delta e^{ i S_{SYM} } }
{\delta V_{mod}(y^{\prime}, \theta,\bar{\theta} ) } U_{\cal C}  \nonumber \\
 &=&  16 g^{2} << \frac{k}{N_{c}}   tr  \frac{\delta }
{\delta V_{mod}(y^{\prime},\theta,\bar{\theta}) } U_{\cal C} >> \;\;\;.
\eeqn
 We have now replaced $y$ by $y^{\prime}$ and saved $y$  for the integration
  variable.
 To evaluate the right hand side,  we compute (again in the $y$ coordinate)
\beq 
\left( \frac{\delta}{\delta V_{mod}(y^{\prime}, \theta,\bar{\theta}) }
 \right)_{i}^{~~j}
 \left( \phi(y, \eta, \bar{\eta}) \right)_{k}^{~~\ell}\;\;\;.
\eeq
 The calculation is long and space permits us to present the final result
 only:
\beqn
\label{eq:ddv}
&& \left( \frac{\delta}{\delta V_{mod}(y^{\prime}, \theta,\bar{\theta}) }
 \right)_{i}^{~~j}
 \left( \phi(y, \eta, \bar{\eta}) \right)_{k}^{~~\ell} \nonumber \\
&& =  (T^{r})_{i}^{~~j} \left \{ \left ( 
-\frac{i}{2} dy^{a} (\eta - \theta) \sigma_{a} (\bar{\eta} - \bar{\theta})
 -\frac{1}{2}\delta (\eta - \theta) dy^{a} \bar{\theta}
 \bar{\sigma}^{b} \sigma_{a} \bar{\eta} {\cal D}_{b}  \right. \right.    \\
&& 
  \left. \left. + \bar{\eta}\bar{\eta} d\eta (\eta-\theta)
 +i \delta(\eta -\theta) \bar{\eta}\bar{\eta} d \eta \sigma^{a} \bar{\theta}
 {\cal D}_{a}
   \right) T^{r}
 \delta^{(4)}(y- y^{\prime}) \right \}_{k}^{~~\ell}   \nonumber   \\
&& +  \left(T^{r} \right)_{i}^{~~j} 
 \frac{i}{4} \delta(\eta-\theta) \bar{\eta} \bar{\eta}
dy^{a} \left( [ \bar{\theta} \bar{\sigma}_{a}W, T^{r} ] \right)_{k}^{~~\ell}
 \delta^{(4)}(y-y^{\prime}) \nonumber \\
&& + \left( \frac{1}{2} dy^{a} \eta \sigma_{a} \bar{\eta} \theta \theta
  +i \theta \theta \bar{\eta}\bar{\eta}d \eta \eta \right)
 \left \{  \left([ T^{r}, \bar{\theta}\bar{\lambda} ] \right)_{i}^{~~j}
 \left(T^{r}\right)_{k}^{~~\ell}  \right.  \nonumber \\
&& \left.  - \left( T^{r} \right)_{i}^{~~j}
 \left([\bar{\theta}\bar{\lambda}, T^{r}] \right)_{k}^{~~\ell}
 \right \} \delta^{(4)}(y-y^{\prime}) \;\;\;. \nonumber
\eeqn
  Here ${\cal D}_{a} \cdots \equiv [ \partial_{a} - \phi_{a}, \cdots]$.
  We have repeatedly used
\beq
\frac{i}{2} ( v_{a} -i \bar{\lambda} \bar{\sigma}_{a} \eta)
 = - (1 + \bar{\eta}^{ \dot{\alpha}} \bar{D}_{\dot{\alpha}} ) \phi_{a}\;\;\;
\eeq
  to convert the expression into  the one containing  the covariant derivative
  with respect to $\phi_{a}$ alone.

Using eq.~(\ref{eq:ddv}) and  the completeness  relation, we are able to
 complete  the right hand side of eq.~(\ref{eq:ssd}). 
 The last line of eq.~(\ref{eq:ddv}) disappears after  the completeness
  relation is used and the covariant derivative
 ${\cal D}_{a}$ and
 $\frac{i}{2}(W\sigma_{a})_{\dot{\alpha} }$  are replaced  respectively by 
 the ordinary derivative $\partial_{a}$ and
  the area derivative $\partial/\partial \Sigma^{a \dot{\alpha}}$. 
  These are  nontrivial properties revealed by our calculation.
 All in all, we find
\beqn
\label{eq:result}
&&\sigma_{\alpha \dot{\beta}}^{a} \epsilon^{\dot{\beta} \dot{\alpha} }
   D^{\alpha} \frac{\partial }{\partial \Sigma^{ a \dot{\alpha}}
  (y^{\prime}) }
  {\bf W}[{\cal C}]  \nonumber \\
&=& 16 g^{2}  \oint \left[ dy^{a} \left( -\frac{i}{2} (\eta-
 \theta) \sigma_{a}(\bar{\eta}- \bar{\theta}) - \frac{1}{2}
  \delta(\eta- \theta) \bar{\theta} \bar{\sigma}^{b} \sigma_{a} \bar{\eta}
  \partial_{b}^{(y)} \right) \right.   \nonumber   \\
&+& \left.  \bar{\eta}\bar{\eta} d\eta (\eta -\theta) + i \delta(\eta -\theta)
 \bar{\eta}\bar{\eta}d \eta \sigma^{a} \bar{\theta} \partial_{a}^{(y)}
 \right]  \nonumber \\
&\times& << \frac{k^{2}}{N_{c}} \left( tr U_{ {\cal C}_{y y^{\prime}} } \right)
\delta^{(4)}(y-y^{\prime})
 \left( tr U_{ {\cal C}_{y^{\prime} y}} \right) >>
 \nonumber \\
&+& \frac{16 g^{2}}{8} \oint  dy^{a}
 \left( \bar{\sigma}^{b} \sigma_{a} \bar{\theta} \right)^{\dot{\alpha}}
 \delta(\eta -\theta) \bar{\eta} \bar{\eta} \nonumber \\
 &\times& << \frac{k^{2}}{N_{c}} \left( tr U_{ {\cal C}_{y y^{\prime}} }
 \right)   \left( \stackrel{\leftrightarrow}{\frac{\partial}
{\partial \Sigma^{b \dot{\alpha}}  } }  \right)
\delta^{(4)}(y-y^{\prime})  \left(tr U_{ {\cal C}_{y^{\prime} y}} \right)>>
 \;\;\;, \nonumber  \\
&\equiv& 16g^{2}N_{c} \oint  dy^{A} << \left(\frac{k}{N_{c}}\right)^{2}
 \left( tr U_{ {\cal C}_{y y^{\prime}} }
 \right)     \hat{ {\cal O}}_{A}   \left( 
\frac{\partial} {\partial \Sigma (y) }, \frac{ \partial}
{\partial y }, y  \right) 
 \nonumber \\
&&      \delta^{(4)}(y-y^{\prime}) 
 \left(tr U_{ {\cal C}_{y^{\prime} y}} \right)>>
\eeqn
  where $A \stackrel{\leftrightarrow}{\frac{\partial}
{\partial \Sigma^{a \dot{\alpha}}  } } B =  A \left(\frac{\partial}
{\partial \Sigma^{a \dot{\alpha}}  } B \right)
   - \left( \frac{\partial}
{\partial \Sigma^{a \dot{\alpha}}  } A  \right) B $.
 It is remarkable that  the right-hand side of this equation  does not have
 any field dependence other than  the Wilson-loop variable in superspace.
Again we can immediately  write down  the rest of the Schwinger-Dyson
  equations:
\beqn
&&\sigma_{\alpha \dot{\beta}}^{a} \epsilon^{\dot{\beta} \dot{\alpha} }
   D^{\alpha} \frac{\partial }{\partial \Sigma^{ a \dot{\alpha}}
  (y^{\prime }) }
  << \left( \frac{k}{N_{c}} \right)^{n}
    \prod_{i=1}^{n} tr U_{ {\cal C}^{(i)} } >>    \nonumber \\
 &=&   16 g^{2} N_{c}
   \sum_{j=1}^{n} \Theta ( y^{\prime } \in {\cal C}^{(j)})
\oint_{ {\cal C}^{(j)} }   dy^{A} <<  \left( \frac{k}{N_{c}} \right)^{n+1}
 \left( \prod_{i=j+1}^{n} tr U_{ {\cal C}^{(i)} } \right)
  \nonumber \\   
&\times&   \left(\tr U_{{\cal C}_{y y^{\prime}}^{(j)}} \right)
   \hat{ {\cal O}}_{A}    \delta^{(4)}(y-y^{\prime})
 \left( tr U_{{\cal C}_{y^{\prime} y }^{(j)}  } \right) 
  \left( \prod_{i=1}^{j-1} tr U_{ {\cal C}^{(i)} } \right) >> \;.
\eeqn
  When  a matter chiral multiplet is  present, we should add   to the right
  hand side  a term
\beqn
&& 16g^{2} 
  \sum_{j=1}^{n} \Theta ( y \in {\cal C}^{(j)})
 <<  \left( \frac{k}{N_{c}} \right)^{n}
 \left( \prod_{i=j+1}^{n} tr U_{ {\cal C}^{(i)} } \right)  \nonumber \\   
&\times& \left(tr J( y^{\prime}, \theta, \bar{\theta})
 U_{{\cal C}_{y^{\prime} y^{\prime} }^{(j)} } \right) 
 \left( \prod_{i=1}^{j-1} tr U_{ {\cal C}^{(i)} } \right) >>  \;.
\eeqn
  Here $J( y^{\prime}, \theta, \bar{\theta}) \equiv i \delta S_{matter}/
 \delta V_{mod}(y^{\prime}, \theta, \bar{\theta})$  and $S_{matter}$
  is $\int d^{4}x~\Phi^{\dagger} e^{V} \Phi \mid_{\theta \theta \bar{\theta}
  \bar{\theta} }$  for fundamental matter and
  $ \int d^{4}x~ tr \Phi^{\dagger} e^{V} \Phi e^{-V}
 \mid_{\theta \theta \bar{\theta}  \bar{\theta} }$  for adjoint matter. 

 In the large $N_{c}$ limit (this time $\lambda_{c} \equiv 4 g^{2}N_{c}$), 
 eq.~(\ref{eq:result})  becomes a closed loop equation for ${\bf W}[{\cal C}]$:
\beqn
\label{eq:sloop} 
&&  \frac{1}{ 4 \lambda_{c}} \sigma_{\alpha \dot{\beta}}^{a}
 \epsilon^{\dot{\beta} \dot{\alpha} }
 D^{\alpha} \frac{\partial }{\partial \Sigma^{ a \dot{\alpha}} }
  {\bf W} [{\cal C}]  \nonumber \\
&=&  \oint \left[ dy^{a} \left( -\frac{i}{2} (\eta-
 \theta) \sigma_{a}(\bar{\eta}- \bar{\theta}) - \frac{1}{2}
  \delta(\eta- \theta) \bar{\theta} \bar{\sigma}^{b} \sigma_{a} \bar{\eta}
  \partial_{b}^{(y)} \right) \right.   \\
 &+& \left. \bar{\eta}\bar{\eta} d\eta (\eta -\theta) + i \delta(\eta -\theta)
 \bar{\eta}\bar{\eta}d \eta \sigma^{a} \bar{\theta} \partial_{a}^{(y)}
 \right] {\bf  W}[{\cal C}_{y y^{\prime}} ]
\delta^{(4)}(y-y^{\prime}){\bf W}[{\cal C}_{y^{\prime} y} ]
 \nonumber \\
 &+&   \frac{1}{8}  \oint  dy^{a}
 \left( \bar{\sigma}^{b} \sigma_{a} \bar{\theta} \right)^{\dot{\alpha}}
 \delta(\eta -\theta) \bar{\eta} \bar{\eta}
{\bf W} [{\cal C}_{y y^{\prime}}]
  \left( \stackrel{\leftrightarrow}{\frac{\partial}
{\partial \Sigma^{b \dot{\alpha}}  } }  \right)
\delta^{(4)}(y-y^{\prime}) {\bf W}[{\cal C}_{y^{\prime} y}]\;\;. \nonumber
\eeqn
  Eqs.~(\ref{eq:result})--(\ref{eq:sloop}) are the results  from
 our discussion.

  Just as in  the non-supersymmetric case, there remain a number of questions
  on how one can manage to extract physics results 
 from eqs.~(\ref{eq:result}), (\ref{eq:sloop}). These may involve the
  problem of renormalizations and  more constructive formulations for instance.
  Originally we were partially motivated by   the recent developments
  associated with the exact evaluation of the light effective action
  beginning with \cite{SW}
  and the surprising connection to integrable systems of particles \cite{Int}.
   A connection of the results presented  in this paper with these is still
 remote but   certainly we are now urged to strive for further
 developments. 

   We thank  Hikaru Kawai for a helpful comment on our approach.

\end{document}